\documentstyle[prd,aps, 
preprint,
multicol,axodraw]{revtex}

\begin{document}

\draft
\preprint{\begin{tabular}{r}
KIAS-P99113 \\
hep-ph/9912524
\end{tabular}}

\title{Radiative Splitting of Three Degenerate Neutrinos}
\author{Eung Jin Chun
and Sin Kyu Kang }
\address{Korea Institute for Advanced Study, 
    Seoul 130-012, Korea \\
{\it Email addresses: ejchun@kias.re.kr, skkang@kias.re.kr}    }
\maketitle

\begin{abstract}
We propose a radiative origin of the two mass splittings of 
three degenerate Majorana neutrinos. 
It can be achieved by extending the standard model to have
the usual effective dimension 
5 operators generating SO(3)-invariant tree-level masses, 
and a charged scalar singlet coupling with the leptons
preserving the U(1) subgroup of SO(3).  
The mass splittings for the atmospheric and solar neutrino oscillations
then arise from one-loop corrections due to the charged scalar singlet 
coupling and the usual tau Yukawa coupling, respectively.
\end{abstract}


\newpage

Current data from atmospheric \cite{atms} and solar \cite{sols}
neutrino observations and terrestrial neutrino experiments 
\cite{lsnd,chooz,heidel} provide meaningful constraints on neutrino 
masses and mixing.  When one takes also into account cosmological indications 
for the existence of hot dark matter \cite{hdm}, 
neutrinos are required to be degenerate in mass \cite{camo}.
To accommodate all of these neutrino data,   at least four neutrinos are
required \cite{sterile}.  If we, however, leave out the not yet confirmed
LSND results \cite{lsnd}, the atmospheric and solar neutrino anomalies
can be explained through neutrino oscillations among three active 
species, $\nu_e, \nu_\mu$ and $\nu_\tau$. 
The atmospheric neutrino oscillation indicates the maximal mixing
between $\nu_{\mu} $ and $\nu_{\tau}$ with a mass squared difference
$\Delta m^2_{atm} \simeq 10^{1.5}~\mbox{eV}^2$\cite{atms}.
The solar neutrino anomaly can be explained through matter enhanced
neutrino oscillation if
$3\times 10^{-6}\leq \Delta m^2_{sol} \leq 10^{-5} ~\mbox{eV}^2$ and
$2\times 10^{-3}\leq \sin^2 2 \theta _{sol}\leq 2\times 10^{-2}$
(small angle MSW), or
$10^{-5}\leq \Delta m^2_{sol} \leq 10^{-4} ~\mbox{eV}^2$,
$\sin^2 2 \theta _{sol}\geq 0.5$ (large angle MSW), $\Delta m^2_{sol}\sim
10^{-7} ~\mbox{eV}^2, \sin^2 2 \theta_{sol}\sim 1.0$ (LOW solution)
\cite{LOW} and through long-distance  vacuum oscillation if
$5\times 10^{-11}\leq \Delta m^2_{sol} \leq 10^{-9} ~\mbox{eV}^2$,
$\sin^2 2 \theta _{sol}\geq 0.6$.
Furthermore, combination of
the cosmological requirement and non-observation of neutrinoless double-beta 
decay \cite{heidel}
singles out a specific pattern of three Majorana neutrino mass matrix 
with almost degenerate mass eigenvalues and bimaximal mixing 
for the atmospheric and solar neutrino oscillations \cite{viss}.
In the leading term, this mass matrix in the charged-lepton flavor basis
is given by 
\begin{equation} \label{bide}
 M^\nu_0 \sim m_0 \pmatrix{ 
    0 & {1\over \sqrt{2}} & {1\over \sqrt{2}} \cr
    {1\over \sqrt{2}} & {1\over 2} &  -{1\over 2} \cr
    {1\over \sqrt{2}} & -{1\over 2} &  {1\over 2} } \,,
\end{equation}
where $m_0 \sim 2$ eV is needed for neutrino hot dark matter.
This brings us to a theoretical challenge to answer the questions:
what is the origin of such a mass pattern?, and 
how can one obtain naturally the desired tiny mass differences?

\medskip

In this letter, we will suggest a simple model in which 
{\it both} atmospheric  {\it and} solar neutrino mass splittings 
are generated from radiative corrections while keeping almost bimaximal
mixing among three active neutrinos.  The degeneracy of 
the three neutrinos would be a consequence of non-Abelian family symmetry,
like SO(3) with three lepton doublets (and right-handed neutrinos) 
transforming as a triplet.  Then, there must be some sector to break
the family symmetry to produce the tiny mass splittings at the 
level of $\Delta m^2_{atm}/m_0^2$ and $\Delta m^2_{sol}/m_0^2$ for 
the atmospheric and solar neutrino oscillations, respectively.
There exist in the literature \cite{degens} several models to explain
both splittings by some textures at tree level  or the solar mass
splitting by loop corrections.    
Our proposal here is to generate both splittings radiatively at one loop
level, and thus we do not require any undesirable fine-tuning of
parameters.  In particular, the finer splitting, $\Delta m^2_{sol}/m_0^2$,
for the solar neutrinos arises from the inevitable one-loop correction
due to the small tau Yukawa coupling \cite{rge}.  For the generation of the 
larger splitting $\Delta m^2_{atm}/m_0^2$,  we introduce a charged Higgs
singlet which couples to the leptonic sector in the same way
appeared in Zee model \cite{zee}

\medskip

Let us first discuss neutrino mass matrix at tree level from which
the degenerate neutrino spectra can be obtained.
Such a mass matrix can be constructed from symmetry principle.
In this work, we impose $SO(3)$ family symmetry for the Majorana neutrino 
sector.
Let $L_i=(\nu_i, l_i)$ be the lepton doublet, where the subscript $i$ refers
to the $(+,-,0)$ component of an $SO(3)$ triplet.
Then, the $SO(3)$ invariant Majorana neutrino mass matrix can come from
the effective dimension five operator,
\begin{equation} \label{Leff}
L_{eff} = {h_\nu \over 2 M} 
        (2 L_+L_- +  L_0 L_0)(\bar{H}\bar{H}) + h.c.\,,
\end{equation}
where $\bar{H}$ is the Higgs doublet coupling to the up-type quarks
and $M\sim 10^{13}$ GeV is the see-saw scale.
Note that the effective operator (\ref{Leff}) arises below the scale $M$
through the see-saw mechanism endowed with heavy right-handed 
neutrinos \cite{seesaw} or a  heavy triplet scalar \cite{triplet}.
Our discussions do not depend on either types of heavy fields at the 
scale $M$.
The effective Majorana neutrino mass matrix in the SO(3) flavor basis is 
\begin{equation}
 M^\nu_0 = \pmatrix{ 0& m_0 & 0 \cr m_0 & 0 & 0 \cr 0 & 0 & m_0 }
\end{equation}
with $m_0= h_\nu \langle \bar{H}^0 \rangle^2/M$.

The SO(3) symmetry has to be broken badly in the charged-lepton Yukawa
sector.  To obtain the neutrino mass matrix (\ref{bide}) in the 
charged-lepton flavor basis, we require that the SO(3) symmetry breaking 
is arranged to yield \cite{ma2}
\begin{equation} \label{LYuk}
 L_{Yuk} = h_\tau (s_1 L_- + c_1 L_0) \tau^c H +\cdots  \,,
\end{equation}
where $c_1=\cos\theta_1$, {\it etc}.
Here we omitted the smaller Yukawa couplings for the first two generations
which are irrelevant for our discussions.
The required SO(3) symmetry breaking in the charged-lepton
sector (\ref{LYuk}) would be obtained by introducing some SO(3) ``flavon''
fields \cite{flavon}.
The lepton doublet fields in the  SO(3) basis is then related to the fields
in the charged-lepton flavor basis as follows;
\begin{equation}
L_{+} = L_e, ~~~~ L_{-} = c_{1}L_\mu - s_{1}L_\tau, ~~~~
L_{0} = s_{1} L_\mu +c_{1} L_\tau .  \,
\end{equation}
This leads to the neutrino mass matrix in the charged-lepton flavor basis,
\begin{eqnarray}
M_{0}^{\nu} =  R_{23}(\theta_{1})
        \cdot \pmatrix{0&m_0&0\cr
               m_0& 0 & 0\cr 0& 0 & m_0\cr}
        \cdot R^{T}_{23}(\theta_{1})  \,.
\end{eqnarray}
Then, we have  the required bimaximal mixing matrix,
\begin{equation} \label{U0}
 U=\pmatrix{ {1\over\sqrt{2}} & {1\over\sqrt{2}} & 0 \cr
              -{c_1\over\sqrt{2}} & {c_1\over\sqrt{2}} & s_1 \cr
              {s_1\over\sqrt{2}} & -{s_1\over\sqrt{2}} & c_1 \cr}
\end{equation}
for $c_1\approx s_1$, and the degenerate mass eigenvalues $(-m_0, m_0, m_0)$.  

\medskip

A degenerate mass pattern at tree level can be modified 
significantly by radiative corrections.
The one-loop corrected neutrino mass matrix due to divergent wave
function renormalization takes the form,
\begin{equation}  \label{loopmm}
M^{\nu} = M^{\nu}_{0} + \frac{1}{2}\left(I\cdot M^{\nu}_0
        + M^{\nu}_0\cdot I\right),
\end{equation}
where $I$ is a matrix of regularized one-loop integrals of neutrino
self-energy diagrams.  
One of the important contribution to the one-loop correction comes from 
the renormalization group evolution below the see-saw scale $M$
thanks to the tau Yukawa coupling  \cite{rge2}.
This gives the nonzero component in the charged-lepton flavor basis,
\begin{equation} \label{eptau}
 I_{\tau\tau} \approx {h_\tau^2 \over 32\pi^2} 
 \ln{M \over M_Z} \equiv \epsilon_\tau\,.
\end{equation}
Including this effect, we get the one-loop corrected mass matrix
in terms of the SO(3) eigenstates,
\begin{eqnarray}
M^{\nu} = m_0 \pmatrix{
 0 & 1 + \frac{1}{2}s_{1}^2\epsilon_{\tau}
  & -\frac{1}{2}c_{1}s_{1}\epsilon_{\tau}  \cr
  1+\frac{1}{2}s_{1}^2\epsilon_{\tau} & 0 &
   -\frac{1}{2}c_{1}s_{1}\epsilon_{\tau} \cr
   -\frac{1}{2}c_{1}s_{1}\epsilon_{\tau}  &
   -\frac{1}{2}c_{1}s_{1}\epsilon_{\tau}  &
   1+c_{1}^2\epsilon_{\tau} \cr}
\end{eqnarray}
where $ \epsilon_{\tau} \approx 10^{-5}$.
Diagonalizing the mass matrix $M^{\nu}$, one finds that 
the mass splittings for the solar and atmospheric neutrino
oscillations are of the same order, that is, $\Delta m_{sol}^2 
\approx \Delta m_{atm}^2   \approx  m_0^2 \epsilon_\tau$.
Therefore, it is impossible to provide the relevant mass splittings for 
both solar and atmospheric neutrinos within this model.

\medskip

To get the correct mass splittings, we need more corrections arising 
from some other flavor violating interactions in the lepton sector.
In this work, we will show that it can be achieved by
introducing a charged scalar singlet $\phi^{+}$, which allows for the
couplings $f_{ij} L_i L_j \phi^+$ \cite{zee}.  Note that these couplings
cannot be SO(3)-invariant due to antisymmetry between lepton doublets, 
that is, $f_{ij}=-f_{ji}$.
The relevant flavor violating  interaction term for our purpose is then
\begin{equation} \label{Ladd}
L_{add}=fL_{+}L_{-}\phi^{+}
\end{equation}
which respects the U(1) subgroup of the SO(3) family symmetry.
Conservation of this U(1) is crucial to maintain the degeneracy between
the first two eigenvalues at the level of the desired degree, as will 
become clear in the following discussions.
Let us recall that there may exist additional finite one-loop corrections in 
Eq.~(\ref{loopmm}) arising  from the interaction term (\ref{Ladd}) 
in the context of two Higgs doublet models \cite{zee2}.
With one Higgs doublet, we do not have these finite corrections.

In Fig.1,  we present the one-loop diagram for neutrino masses
generated from the above Lagrangian (\ref{Ladd}).
The resulting one-loop integrals are  
\begin{equation} \label{epf}
I_{++} = I_{--} \approx  \frac{f^2}{32\pi^2} 
     \ln{M_{\phi^+} \over M_Z} \equiv \epsilon_f.
\end{equation}
Then, the one-loop corrected neutrino mass matrix becomes
\begin{eqnarray} \label{finalmm}
M^{\nu} = m_0 \pmatrix{ 0 & 1+\epsilon_f + \frac{1}{2}s_{1}^2
          \epsilon_{\tau} & -\frac{1}{2}c_{1}s_{1} \epsilon_{\tau} \cr
 1+\epsilon_f + \frac{1}{2}s_{1}^2  \epsilon_\tau
         & 0 &-\frac{1}{2}c_{1}s_{1}\epsilon_{\tau} \cr
         -\frac{1}{2}c_{1}s_{1}\epsilon_{\tau} &
         -\frac{1}{2}c_{1}s_{1}\epsilon_{\tau} &
           1+c_{1}^2\epsilon_{\tau} \cr}.
\end{eqnarray}
In the leading order, the mass eigenvalues are
\begin{eqnarray}
 m_1^2 &=& m_0^2(1+\epsilon_f + \frac{1}{2}s_{1}^2\epsilon_{\tau})^2 
          \nonumber\\
 m_2^2 &=& m_0^2(1+\epsilon_f + \frac{1}{2}s_{1}^2\epsilon_{\tau}
          +\frac{s_{1}^2 c_{1}^2}{2}\frac{\epsilon_{\tau}^2}
           {\epsilon_f})^2 \nonumber\\
 m_3^2 &=& m_0^2(1+ c_{1}^2\epsilon_{\tau})^2 \,.
\end{eqnarray}
The atmospheric and solar neutrino mass-squared differences 
are then given by
\begin{eqnarray} \label{twoD}
\Delta m^2_{atm} &=& \Delta m_{32}^{2} \approx  2m_0^2\,\epsilon_f   \nonumber\\
\Delta m^2_{sol} &=& \Delta m_{21}^{2} \approx {1\over4} m_0^2 \sin^22\theta_1
       {\epsilon^2_{\tau} \over \epsilon_f}.
\end{eqnarray}
For $m_0 \sim 2$ eV, we get the right value of mass splitting 
for the atmospheric neutrinos with $\epsilon_f \sim  10^{-3}$.
Let us remark that if one introduces the terms which break the U(1) subgroup
of the SO(3) family symmetry like $f^{\prime}L_{\pm}L_0 \phi^+$, it will produce
a too large splitting, $\Delta m^2_{21} \sim m_0^2 \epsilon_{f^{\prime}}$,
unless the coupling $f^{\prime}$ is suppressed enough.
The relation in Eq.~(\ref{twoD}) reproduces the simple connection 
between the atmospheric and solar neutrino oscillations \cite{ma2}
\begin{equation} \label{REL}
 {\Delta m^2_{atm} \Delta m^2_{sol} \over m_0^4 \sin^22\theta_1}
 \approx {1\over2} \epsilon_\tau^2
\end{equation}
without resorting to {\it ad hoc} tree-level splitting between
the first two and third neutrino masses.  From the above relation, it
turns out that our model picks out the MSW solution with 
lower mass-squared difference, $\Delta m^2_{sol} \sim 10^{-7}$ eV$^2$,
which is often disregarded in discussions.
Contrary to the conclusion in Ref.~\cite{ma2}, it is rather hard to get
the vacuum oscillation solution to the solar neutrino problem
due to the $\epsilon_\tau$ effect with logarithmic enhancement.

\medskip

There is more freedom in the two Higgs doublet model. 
In this case, the expression for $\epsilon_\tau$ (\ref{eptau})
contains the additional factor $2\tan^2\beta$ where $\tan\beta$ is 
the ratio between the vacuum expectation values of two Higgs fields.
Therefore, the relation (\ref{REL}) is modified to
\begin{equation} \label{REL2}
 {\Delta m^2_{atm} \Delta m^2_{sol} \over m_0^4 \sin^22\theta_1}
 \sim 10^{-10}\tan^4\beta\,.
\end{equation}
For $m_0 \sim 2$ eV and $\tan\beta \sim 3$, we get $\Delta m^2_{sol}
\sim 2\times10^{-5}$ eV$^2$ which is in the right range for the  
large mixing angle MSW solution.
With two Higgs doublets, there could arise large finite one-loop masses 
through Zee mechanism \cite{zee,zee2} in the presence of the coupling 
$\mu H_1 \bar{H}_2 \phi^+$.  This can be suppressed when
the $\mu$ term is absent, or the charged scalar singlet has a mass at the
see-saw scale, $M_{\phi^+} \sim M$.

\medskip

Let us finally consider the change of mixing angles from 
their tree level values in Eq.~(\ref{U0}) due to the radiative
corrections.
The neutrino mixing matrix $U$ which is obtained from re-diagonalization 
of the one-loop corrected mass matrix (\ref{loopmm}) can be parameterized by
\begin{equation}
U= R_{23}(\theta_1+\delta \theta_1) 
   \cdot R_{13}(\theta_2+ \delta\theta_2) 
   \cdot R_{12}(\theta_3+\delta\theta_3) 
\end{equation}
where $\theta_1 \approx \pi/4$, $\theta_2=0$ and 
$\theta_3=\pi/4$ coming from Eq.~(\ref{U0}). 
The mass pattern (\ref{finalmm}) leads to the vanishing corrections
$\delta\theta_{2,3}$ as long as the $\mu$ and $e$ 
Yukawa couplings are neglected.   Furthermore,
thanks to the hierarchy of $\epsilon_f$ and $\epsilon_{\tau}$,
the angle $\delta \theta_1$ comes out to be as small as $\delta \theta_1 \sim
\epsilon_{\tau}/\epsilon_f \sim \sqrt{\Delta m^2_{sol}/\Delta m^2_{atm}}$, 
which can be estimated by the see-saw diagonalization of (23)-submatrix
of $R_{12}M^{\nu}R_{12}^{T}$.
Thus, the neutrino mixing matrix is quite stable against the above quantum 
corrections.

\medskip

In conclusion,  we presented a simple way to understand the tiny mass
splittings of three degenerate Majorana neutrinos which are good
candidates for hot dark matter of the universe.
Our proposal is to extend the standard model by introducing two additional 
sectors.  One sector consists of the usual effective dimension 
5 operators for tree-level neutrino masses arising from the see-saw mechanism.
Here we impose the non-Abelian family symmetry, SO(3), to enforce 
the degeneracy of three neutrino species.  The other sector contains a
charged scalar singlet coupling with the lepton doublets which breaks SO(3)
but preserves its U(1) subgroup.  
As a consequence, the mass splitting for the atmospheric
neutrino oscillation is generated through (the U(1) preserving)
one-loop corrections with a coupling of ${\cal O}$(0.1).  The tiny 
mass splitting for the solar neutrino oscillation arises then from the
renormalization group effect due to the usual tau Yukawa coupling.
In the case of one Higgs doublet, or two Higgs doublets 
with $\tan\beta\sim 1$, the lower mass-squared differences of the MSW 
solution can be realized.  The large mixing angle MSW solution can
be obtained only for two Higgs doublets with $\tan\beta\sim3$.
It also turns out that the vacuum oscillation solution is disfavored 
in our scheme.

\vspace{2cm}

\begin{center}
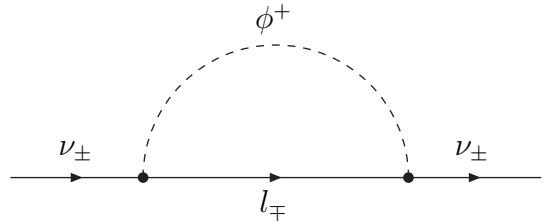
FIG.~1. Neutrino self-energy diagram coming from the charged singlet 
interactions.
\begin{picture}(500,100)(0,0)
%
\Vertex(150,10){2}
\Vertex(250,10){2}
\ArrowLine(100,10)(150,10)
\Text(125,20)[]{$\nu_{\pm}$}
\ArrowLine(150,10)(250,10)
\Text(200,0)[]{$l_{\mp}$}
\ArrowLine(250,10)(300,10)
\Text(275,20)[]{$\nu_{\pm}$}
\DashCArc(200,10)(50,0,180){3}
\Text(200,70)[]{$\phi^+$}
\end{picture}
\end{center}

\end{document}